# Experimental evidence for competition of antiferromagnetic and ferromagnetic correlations in HgCr$_2$S$_4$

V. Tsurkan[1,2], J. Hemberger[1], A. Krimmel[1], H.-A. Krug von Nidda[1], P. Lunkenheimer[1],

S. Weber[1], V. Zestrea[2], and A. Loidl[1]

[1] *Experimental Physics V, Centre for Electronic Correlations and Magnetism, University of Augsburg, D- 86159 Augsburg, Germany*
[2]*Institute of Applied Physics, Academy of Sciences of Moldova, MD-2028, Chişinău, R. Moldova*

## ABSTRACT

A detailed study of poly- and single crystalline samples of the normal spinel HgCr$_2$S$_4$ is reported. The structural refinement reveals enhanced values of the atomic displacements suggesting closeness to a structural instability. Magnetization, electron-spin resonance and specific-heat studies document strong ferromagnetic fluctuations close to 50 K and the occurrence of complex antiferromagnetic order at $T_N$ = 22 K. We found highly unconventional behavior resembling properties of a non-collinear antiferromagnet and soft ferromagnet dependent on temperature and magnetic field. Even weak external magnetic fields disturb the antiferromagnetic order and strongly enhance the ferromagnetic correlations. The conductivity shows an insulating character with non-monotonous temperature behavior and a metal-to-insulator like transition at 80 K well above the magnetic transition. The observed anomalies are related to bond frustration due to competing magnetic exchange interactions between the Cr ions.

## I. INTRODUCTION

Magnetic compounds with frustrated exchange due to geometric constraints became the subject of current theoretical and experimental interest. Transition-metal oxides and chalcogenides with the general formula AB$_2$X$_4$ exhibiting the spinel structure received special attention. In AB$_2$X$_4$ spinels both cation sites, the tetrahedrally coordinated A-site and the octahedrally coordinated B-site, are geometrically frustrated. The A-sites form two interpenetrating *fcc* lattices, while the B-sites constitute a pyrochlore lattice with corner sharing tetrahedra. In addition another type of frustration governs many magnetic spinel compounds which so far has not been studied in detail. Depending on the lattice constant, nearest neighbor (NN) direct antiferromagnetic (AFM) B-B exchange and indirect NN ferromagnetic 90° (FM) B-X-B superexchange via anions are competing in the oxide spinels.[1] In the chalcogenide spinels, the direct AFM exchange is weakened by the larger inter-cation distance, but more-distant neighbour antiferromagnetic superexchange interactions like B-X-A-X-B or B-X-X-B become important.[2-4] There are cases when the competing interactions are equal and almost cancel each other, revealing a Curie-Weiss temperature close to zero, a situation which one can call bond frustration. So far it is unclear, whether geometrical frustration and bond frustration drive similar exotic ground states. It is common believe that the novel phenomena found recently in magnetic spinels, e.g. heavy-fermion-like liquid,[5-7] spin-singlet states,[8,9] composite spin degree of freedom,[10] spin-orbital liquid and orbital glass states,[11-13] as well as the colossal magneto-capacitive coupling in spinel multiferroics,[14] bear the sign of the geometrical frustration that enhances the spin, orbital and lattice fluctuations. Besides the pure academic interest, this fascinating physics offers new possibilities for spintronic applications and design of devices with advanced functionality.

Most of chalcogenide chromium spinels $ACr_2X_4$ (with A=Cd, Hg; X=S, Se) are simple Heisenberg ferromagnets,[15] and hence are not expected to be geometrically frustrated. However, $HgCr_2S_4$ shows a non-collinear antiferromagnetic order despite a high positive Curie-Weiss temperature of 142 K dominated by strong ferromagnetic exchange.[3] Neutron-diffraction investigations at low temperatures and in zero magnetic field revealed a spiral spin configuration.[16] Moderate fields of only several kilo-oersteds destroy the spiral structure and align the spins ferromagnetically. Early magnetization measurements report a Curie temperature of 36 K,[4] but disagree with the results of neutron diffraction[16] and optical[17] studies which conclude that $HgCr_2S_4$ is an antiferromagnet below a critical temperature of 60 K. The reason of this discrepancy remained unclear.

The scope of our study was to reveal the magnetic, electronic and thermal correlations in a compound with chromium ions occupying solely B-sites of the intrinsically frustrated pyrochlore lattice. This lattice is known to preclude a simple AFM spin arrangement.[18,19] Indeed, $HgCr_2S_4$ manifests a complex antiferromagnetic ground state, however, the ferromagnetic exchange interactions are strong and dominating. Hence, we expect that bond frustration is of utmost importance here. To get information on local and macroscopic magnetic properties we used complementary techniques of DC magnetization, AC susceptibility, electron-spin resonance (ESR), and specific heat.

## II. EXPERIMENTAL DETAILS

Polycrystalline $HgCr_2S_4$ samples were prepared by solid-state reaction from high purity (99, 999%) binary mercury sulfide and elementary Cr and S in evacuated quartz ampoules. The synthesis was repeated several times in order to reach good homogeneity and to minimize non-reacted binary sulfides. The final heat treatment was performed at 800 $^o$C in an atmosphere of sulfur excess. Thus prepared polycrystals were used as starting material for the single-crystal growth. Single crystals were obtained by chemical transport reaction using chlorine as transport agent. The growth experiments were performed at temperatures between 850 and 800 $^o$C. The crystals were obtained in form of perfect octahedra with shiny surfaces and dimensions up to 2 mm on the edge.

Powder x-ray diffraction was performed utilizing Cu-K$_\alpha$ radiation with a wave length $\lambda = 1.5406$ Å and a position-sensitive detector. The magnetization measurements were done utilizing a commercial superconducting quantum interference device (SQUID) magnetometer (MPMS-5, Quantum Design) in fields up to 50 kOe. The heat capacity was measured using a physical properties measurement system (PPMS, Quantum Design) in the temperature range 2 K < T < 300 K and magnetic fields up to 70 kOe. The ESR measurements were performed with a Bruker ELEXSYS E500 CW-spectrometer at X-band frequency ($\upsilon$ = 9.36 GHz), equipped with a continuous gas-flow cryostat for Helium (Oxford Instruments) covering a temperature range between 4.2 K and room temperature. For the ESR experiments thin disks with the faces along the crystallographic (110)-plane were cut from the single crystals. The disks were fixed on the sample holder allowing for rotation of the static magnetic field within the plane of the disk, where the cubic <001>, <110>, and <111> directions can be adjusted. The microwave field was always applied perpendicular to the plane. The resistivity measurements were conducted in the temperature range 5 K < T < 300 K using both two- and four-probe methods.

## III. EXPERIMENTAL RESULTS AND DISCUSSION

### A. Structural analysis

Structural details of powdered single crystals and polycrystals were investigated by conventional x-ray powder diffraction at room temperature. Measured diffraction profiles and refinements for the corresponding samples are shown in Fig.1. The data were analyzed by standard Rietveld refinement using the FULLPROF program.[20] $HgCr_2S_4$ crystallizes in the normal cubic spinel structure with the space group $Fd\bar{3}m$ (No. 227). No indications of impurity phases could be detected. In total 12 parameters have been refined: scale factor, zero-point shift, 3 resolution parameters, lattice parameter $a_0$, sulfur positional parameter $x$ in fractional coordinates (*f.c.*), Hg site-occupation factor (SOF), Cr site-occupation factor, and three isotropic temperature factors, $B_{iso}$, for Hg, Cr and S, respectively. The refined structural parameters are presented in Table 1.

Table 1. Crystallographic data on $HgCr_2S_4$ crystals

| Sample | $a_0$ (Å) | $x$(S) (*f.c.*) | SOF(Hg) | SOF(Cr) | $B_{iso}$(Hg/Cr/S) (Å$^2$) | $R_{Bragg}$ |
|--------|-----------|-----------------|---------|---------|----------------------------|-------------|
| SC | 10.256(1) | 0.267(1) | 1.08(3) | 1.02(3) | 2.2(1)/1.8(2)/2.3(3) | 5.7 % |
| PC | 10.252(1) | 0.265(1) | 1.05(3) | 1.07(3) | 2.3(1)/2.5(2)/1.8(3) | 4.3 % |

The refined structural parameters of single crystals and polycrystalline samples are in very good mutual agreement. As shown in the inset of Fig. 1, there is a slight redistribution of the intensity between the peaks (511) at $2\Theta \approx 46.0^\circ$ and (440) at $2\Theta \approx 50.4^\circ$ for poly- and single crystals, which indicates some difference in the cationic defect structure. Most importantly, large isotropic temperature factors ($B_{iso}=8\pi^2u^2$ with $u$ being a mean quadratic displacement) have been found for all three constituent atoms, which are enhanced by almost an order of magnitude as compared to the vast majority of other magnetic spinel compounds.[21] Such strongly enhanced values of the atomic displacement together with high value of the sulfur positional parameter suggest that $HgCr_2S_4$ is close to a structural instability.

**B. Magnetization.**

Fig. 2 presents the temperature dependences of the magnetization M for poly- and single crystalline samples of $HgCr_2S_4$ measured at different fields. At low fields (H ≤ 1 kOe) the magnetization shows a sharp increase at temperatures below 60 K apparently indicating the development of strong ferromagnetic correlations in agreement with the earlier results.[3] With further decreasing temperatures the magnetization shows a maximum at about 30 K followed by a downturn resembling a transition into an antiferromagnetic state. The maximum in M is shifted to lower temperatures with increasing magnetic fields indicating the suppression of the AFM state. It finally disappears in fields above 10 kOe, where the magnetization monotonously increases with decreasing temperature. We note a strong enhancement of the ferromagnetic correlations by the magnetic field. A standard determination of the transition temperature as routinely performed for ferromagnets by plotting (dM/dT)/M *vs.* T yields a value of about 50 K for measurements in a field of 100 Oe. However we will show below that this is not a temperature that marks the onset of long-range ferromagnetic order but indicates the appearance of strong ferromagnetic correlations. For an external field of 50 kOe, transition temperature increases up to 100 K! Although in the latter case it strictly cannot be defined, such a strong shift of the magnetization to high temperatures by a magnetic field is probably the largest one ever reported in (ferro)magnetic compounds. Already at this point we note that strong ferromagnetic correlations are present in $HgCr_2S_4$ at about 50 K, although the ground state is a complex antiferromagnet. Similar observations have been reported in doped manganites and were interpreted as driven by the competition of charge, spin and orbital degrees of freedom. But we would like to recall that in $HgCr_2S_4$ the *3d* electrons of the Cr ions are rather orbitally inactive due to the half-filled $t_{2g}$ shell. Beside this, all Cr ions are in a 3+ state. Thus, the variation of the magnetic state in $HgCr_2S_4$ must be fully related to the spin degree of freedom.

In the inset of Fig. 2 the temperature dependence of the reciprocal susceptibility $\chi^{-1}$ measured in a field of 10 kOe is plotted both for poly- and single crystalline samples. A linear fit to the experimental data at high temperatures yields a Curie-Weiss temperature $\Theta_{CW} = 140$ K and an effective magnetic moment $p = 3.90$ $\mu_B$. The latter value agrees well with the theoretical spin-only value of 3.87 $\mu_B$ for $Cr^{3+}$ resulting from a $3d^3$ state. The large positive value of $\Theta_{CW}$ reflects the dominance of ferromagnetic interactions in the magnetic exchange. $HgCr_2S_4$ seems to be a unique magnetic sulfo-spinel characterized by such a large positive Curie-Weiss temperature but reveals long-range antiferromagnetic ordering at low temperatures.

Fig. 3a shows the field dependent magnetization M at several temperatures for single crystalline $HgCr_2S_4$ measured along the <001> cube edge direction. According to neutron diffraction the propagation vector of the magnetic spiral is directed along one of the cube edges in a given domain.[16] On decreasing temperature in the range 30 K<T<60 K, the magnetization shows a ferromagnetic-like behavior with a linear increase at low fields due to demagnetization effect followed by saturation in fields above the demagnetization field. Below the AFM transition at $T_N$=22 K (deduced from the specific heat data, see section III.D), the magnetization after zero field cooling starts to reveal a substantial non-linearity in low fields as shown in detail in Fig. 3b. After initial linear growth up to a certain field $H_1$ (1.7 kOe at 2 K), M shows a steeper increase in fields up to $H_2$ (4.5 kOe), continues with a somewhat weaker slope and approaches saturation at the full value of 6 $\mu_B$ as expected for $2Cr^{3+}$ ions (per formula unit) in fields above $H_3$ (8 kOe). On decreasing field a hysteretic behavior appears below $H_2$ where the magnetization continues the linear decrease down to zero field without any anomaly at $H_1$. Note that on reversing the field and repeating the full cycle the hysteresis becomes less pronounced as compared to the virgin curve. The observed hysteresis can be explained by the reorientation of the three possible domains as evidenced by neutron scattering.[16] No remnant magnetization was detected at all temperatures proving the absence of a ferromagnetic component of the spiral in zero field. The strong increase of the saturation field $H_3$ observed below 22 K cannot be attributed to demagnetization field which is of order of 1.8 kOe for the given sample shape. The overall behavior of the magnetization measured along the other direction, <111> is very similar to that observed along the <001> axis, suggesting also a very small magnetocrystalline anisotropy, not enough to explain the high saturation field. Thus, this rather has to be related to the competing exchange interactions on the gradual suppression of the spiral antiferromagnetic order.

Although in the temperature range 30-60 K the typical ferromagnetic correlations are evident, the question arises whether they are connected with spontaneous ferromagnetic order. To check this we performed a scaling analysis of the magnetization data using a modified Arrott plots technique. We probed different critical exponents including mean field and 3D Heisenberg values. A substantial non-linearity of the Arrott plots was present below 10 kOe in contrast to linear behavior at high fields expected for proper ferromagnets. The non-linearity of the Arrott plots probably has to be related again to the competition of the exchange interactions, as we have seen that at low temperatures the same field is necessary to suppress the non-collinear spin arrangement. Thus, the presence of spontaneous magnetization in $HgCr_2S_4$ cannot be confirmed on the basis of magnetization experiments. From the Arrott analysis only an effective Curie temperature of about 50 K can be inferred in agreement with the abovementioned low-field estimate. But obviously no conventional long-range FM state is established below this temperature.

### C. AC susceptibility

The temperature dependence of the real part of the susceptibility $\chi'$ for single crystalline $HgCr_2S_4$, as measured at a driving AC field of 4 Oe at different external DC magnetic fields, is presented in Fig. 4. In the absence of an external magnetic field, $\chi'$ shows a broad maximum located at 30 K similar to the low field DC susceptibility plotted in the same figure. The steepest slope of the left wing of $\chi' = f(T)$ in zero field occurs at 22 K and marks $T_N$. Application of a DC magnetic field strongly suppresses the maximum of $\chi'$ concomitantly shifting it to higher temperatures (right scale of Fig. 4). In the frequency range 1- 1000 Hz the real part of the AC susceptibility is frequency independent. The imaginary part $\chi''$ is about two orders of magnitude smaller than the real part and, hence, practically undetectable due to the uncertainty in phase adjustment. The behavior of the AC susceptibility for the polycrystalline sample looks very similar to that of the single crystals.

### D. Specific heat

Fig. 5 shows the temperature dependence of the specific heat plotted as $C_p/T$ *vs.* T both for poly- and single crystalline $HgCr_2S_4$. In the single crystal the specific heat manifests a well pronounced anomaly at a temperature of 22 K (see inset of Fig. 5). This temperature coincides with the temperature $T_N$ as determined from the AC susceptibility and is far from 50 K estimated from the DC magnetization. As documented in the inset, application of magnetic fields shifts the anomaly in the specific heat to lower temperatures indicating antiferromagnetic correlations. Therefore, it can be associated with the transition into the AFM state. However, the change of the Néel temperature $T_N$ in $HgCr_2S_4$, as evidenced by the specific heat, is of several orders of magnitude higher than for conventional antiferromagnets.[22-24] Indeed, a field of only 1 kOe yields a huge shift of the anomaly in $C_p$ down to 18 K. Concomitantly, a strong broadening of the anomaly at $T_N$ occurs in contrast to the sharp anomaly seen in conventional antiferromagnets. In fields above 5 kOe the anomaly at $T_N$ is completely suppressed. The entropy involved in the transition calculated by integrating the difference $(C_0-C_{5\,kOe})/T$ over the transition region is 0.35 J mol$^{-1}$K$^{-1}$. This value is surprisingly low being only about 1.5 % of the full entropy of 2Rln4 (or 23.12 J mol$^{-1}$K$^{-1}$) expected for the full ferromagnetic alignment of the Cr spins. We additionally note that the magnetic entropy in $HgCr_2S_4$ at the AFM transition is nearly two times lower than that of conventional antiferromagnets where it usually reaches 0.3-0.4 of the full value.

The evolution of the specific heat at higher fields is presented in Fig. 6. At temperatures above 55 K the specific heat in finite fields becomes higher than that in zero magnetic field (Fig. 6a). The difference between $(C_0-C_{10\,kOe})/T$ is shown on an enlarged scale in Fig. 6b for both poly- and single crystals indicating that some weight of the specific heat in the field is shifted to higher temperatures as usually observed in ferromagnets. Fig. 6b also documents that the heat capacity anomaly at $T_N$ in polycrystals is found at slightly higher temperature compared to single crystals. Another feature becomes pronounced at temperatures below 9 K. In fields up to 20 kOe the specific heat is still higher than in the absence of the field (see inset of Fig. 6a). Together with the magnetization data in Fig. 2 this indicates an intermediate regime of competing AFM an FM correlations, when the AFM order is already broken while the FM order is not yet fully established. Only for fields above 30 kOe the situation is reversed and the specific heat decreases as one expects for an ordinary ferromagnet, where the spin-wave dispersion opens a gap proportional to the external field. Thus, below 9 K the ferromagnetic correlations in $HgCr_2S_4$ become dominant only at high fields. To estimate the magnon contribution to the specific heat we used plots $C/T^{3/2}=\delta+\beta_D T^{3/2}$, which for ferromagnets should present a straight line with slope $\beta_D$ and intercept $\delta$ (see inset of Fig. 6 a). The intercept $\delta$ characterizes the magnon contributions and is inversely proportional to the stiffness of the magnon dispersion. The slope $\beta_D$ is determined by the phonon part of the low-temperature specific heat and allows a direct estimate of the Debye temperature. From $\beta_D$ a Debye temperature $\theta_D = 240$ K $\pm$ 5 K was derived. The intercept of the linear high-temperature fits, as shown by the dashed line in the inset of Fig. 6, yields 0.052 J/mol $K^{5/2}$ in zero external field and decreases with increasing fields, resulting in a value of approximately 0.04 J/mol $K^{5/2}$ for 70 kOe. This evidences the increase of the magnon stiffness on increasing external magnetic fields. Moreover, we have to mention the excess specific heat below 7 K at fields smaller than 50 kOe. It cannot be attributed to nuclear Zeeman contributions, because those should be enhanced in a magnetic field. In general, such an excess specific heat is typical for glassy states. It suggests some degree of disorder and frustration in the antiferromagnetic ground state of $HgCr_2S_4$ and intermediate state as well. However, field-cooled and zero field cooled susceptibility experiments do not reveal any irreversibility which would prove some freezing process related to structural disorder. Probably strong spin fluctuations connected with the competition of FM and AFM interactions are responsible for the excess specific heat in low fields. For fields above 50 kOe the energy gap of the magnon excitations becomes evident and increases with the field as theoretically expected.

### E. Electron-spin resonance

Fig. 7 shows the temperature evolution of the ESR spectra for the static magnetic field applied along the <001> axis. Fig. 8 illustrates the corresponding resonance field and linewidth. Starting at high temperatures in the paramagnetic regime, we observe a single exchange-narrowed resonance line without any orientation dependence. It can be well described by a Lorentzian shape. Its resonance field is found at 3366 Oe corresponding to a g-value g = 1.98 as typically found for $Cr^{3+}$.[25] The linewidth (lower frame of Fig. 8), which is of the order of 100 Oe, first decreases on decreasing temperature, reaches a minimum of 77 Oe at 135 K and increases again up to a local maximum of 156 Oe at 60 K, which can be ascribed to critical fluctuations on approaching the ferromagnetic-like transition. Note that the ESR experiments are performed in finite magnetic fields of approximately 3 kOe (see Fig. 7) which are strong enough to destroy long-range antiferromagnetic order (see Fig. 2). Below 60 K the resonance field is already remarkably shifted as compared to high temperatures, but remains still isotropic for rotation of the static field within the plane of the disk. This shift is due to demagnetization fields originating from the increasing magnetization in the disk shaped sample magnetized in plane.[26] On further decreasing temperature the shift saturates near 30 K at 2.2 kOe and additional magnetostatic modes become excited, which are visible as a couple of weaker lines on the right hand wing of the main resonance which is still well described by a Lorentz curve. The existence of the magnetostatic modes corroborates the presence of strong ferromagnetic correlations. The width of the main resonance exhibits another minimum of 63 Oe at 38 K, due to suppression of the ferromagnetic fluctuations, but then strongly broadens indicating spin fluctuations related to the transition into the non-collinear state. At 22 K a second weak resonance like feature appears at low fields (marked by an arrow in Fig. 7) which on cooling continuously shifts to higher fields. Concomitantly the main resonance looses its Loretzian shape and becomes heavily distorted, when merging with magnetostatic modes (T=18 K). Then both resonances overlap (T=13 K) and still deviate from Lorentzian shape. But finally they separate again into two independent broad lines, one Lorentzian at low resonance field (1.5 kOe) and the second with irregular substructure at higher fields. At the same time the spectrum starts to reveal cubic anisotropy. Comparison with our magnetization data allows for identifying the small resonance-like feature which appears at 22 K with the transition into the non-collinear low-field magnetic phase. The reorientation of the magnetization at a certain field gives rise to an ESR signal analogously to the spin-flop transition in conventional antiferromagnets. As long as this spin-flop like transition is at lower fields than the main resonance, this main resonance is characterized as a ferromagnetic resonance because it originates entirely from the phase with ferromagnetic correlations. On decreasing temperature the transition crosses the main resonance, i.e. at low fields the resonance features of the non-collinear phase appear, whereas at high fields the signal has still the ferromagnetic character. In this temperature regime the spectrum is difficult to describe by fit curve, and hence it is marked by the dashed area in Fig. 8. At low temperature below the crossover only the resonance of the non-collinear phase remains together with the spin-reorientation signal at high fields. Substructure and quite large width of the high-field signal reflect the gradual change of the magnetization from the spiral into the ferromagnetic structure found both in the specific heat and magnetization.

## IV. CONCLUDING REMARKS

Our studies of the frustrated antiferromagnetic spinel $HgCr_2S_4$ reveal unconventional behavior of the magnetic properties and specific-heat. The ground state is extremely sensitive to external fields. The non-collinear antiferromagnetic spin configuration established below 22 K is suppressed by fields as low as 8 kOe. Moreover, the anomaly in the specific heat related to the antiferromagnetic transition is characterized by very low entropy of the order of 1 % of the full magnetic entropy. The magnetization and ESR measurements clearly evidence the appearance of strong ferromagnetic correlations below 50 K. At the same time, no conventional long-range ferromagnetic order sets in and respectively, no anomaly in the specific heat at around the supposed ferromagnetic transition is present. Similarly, we have to mention that in the conductivity no anomaly was found at about 50 K, but it exhibits a very unusual temperature dependence contrasting to that observed in other related chromium chalcogenide spinels. It manifests an almost logarithmic decrease down to 80 K followed by a metal-to-insulator (MI) like transition with a strong increase by almost 8 decades reaching a maximum at 25 K. Roughly at $T_N$ this metal-like behavior changes to a semiconducting behavior again. The reason of the strong increase of the conductivity below 80 K remains unclear. We would like to recall that the ferromagnetic correlations start to dominate the magnetization at about 50 K, i.e. approximately in the middle of the metal-like temperature regime which suggests some relation between conductivity and ferromagnetic correlations. This assumption is supported by the fact that the application of the magnetic field shifts the MI transition to higher temperatures which strongly correlates with the shift of magnetization. This implies an extremely high magnetoresistance effect which reaches 6 orders of magnitude.[27] A metal-to-insulator like transition and strong magnetoresistance effect in the related spinel $FeCr_2S_4$ have been explained by spin-disorder scattering.[28-30] We think that this explanation cannot be valid in $HgCr_2S_4$ due to the fact that in the "metallic" very broad regime between 80 K and 25 K, strong ferromagnetic fluctuations are established, but subsequently become suppressed on approaching $T_N$ at 22 K.

Keeping in mind the peculiar arrangement of the Cr ions forming the corner-sharing tetrahedral network one may attribute the anomalous properties of $HgCr_2S_4$ to geometrical frustration. However, the frustration parameter $f$ (defined as the ratio of the Curie-Weiss temperature $\Theta_{CW}$ to the Néel temperature $T_N$) of about 6 for $HgCr_2S_4$ is below the limit of 10 usually taken to distinguish strongly geometrically frustrated magnets.[19] The frustration parameter of the related sulfide spinel $ZnCr_2S_4$ is even lower (about 1), but it similarly develops a complex AFM spin arrangement at low temperatures apparently connected with strong magnetic frustration.[31] The geometrical frustration dominates in the oxide spinels, e.g., $ZnCr_2O_4$ and $CdCr_2O_4$, where it is linked to strong direct AFM exchange.[32,33] It seems to be less effective in the AF sulfide spinels. As we already mentioned, in $HgCr_2S_4$ the direct AFM exchange is considerably reduced in favor of large ferromagnetic NN superexchange due to the larger inter-atomic distance and results from the positive Curie-Weiss temperature of $HgCr_2S_4$ ($\Theta_{CW} = +140$ K) compared to $ZnCr_2O_4$ ($\Theta_{CW} = -390$ K). Our results suggest that it is the bond frustration due to competing NN and NNN interactions that dominates the properties of $HgCr_2S_4$. It is important to note that in contrast to $HgCr_2S_4$, the other sulfide spinel $CdCr_2S_4$ shows only pure ferromagnetic properties below the ordering temperature of 85 K, although it has just the same distance between the magnetic cations as $HgCr_2S_4$ and only slightly larger Curie-Weiss temperature ($\Theta_{CW} = 152$ K). At first sight it is quite surprising that this insignificant difference in the relative strength of competing interactions might be responsible for the high frustration in $HgCr_2S_4$. But the difference in the ferromagnetic and antiferromagnetic exchange constants for these two compounds estimated in Ref. 3 within mean-field approximation from the difference in their Curie-Weiss temperatures yields a value of about 1 K for FM and 0.3 K for AFM exchange which corresponds well to the Zeeman energy necessary to suppress the spiral structure in $HgCr_2S_4$. Thus far, a self-consistent model of the exchange interactions in this group of magnetic compounds is absent. The other type of interactions beside the direct exchange and superexchange might be also important, for example, a biquadratic exchange as proposed by Grimes and Isaac.[34] Enhanced values of the sulfur positional parameter together with very high value of the temperature factors in $HgCr_2S_4$ (see Table 1) suggest a possible off-centre displacement of the octahedral Cr ions. In this case the biquadratic exchange can be considerably enhanced at low temperatures.[34] It can additionally contribute to bond frustration that is to our opinion responsible for the unusual ground state in $HgCr_2S_4$.

In summary, structural, magnetization, electron-spin resonance, conductivity, and specific heat studies have been performed on $HgCr_2S_4$ poly- and single crystals. The high values of the atomic displacements suggest proximity to a structural instability. Ferromagnetic correlations appear below 60 K and are strongly enhanced by magnetic field. However, no conventional long-range ferromagnetic order is established. The anomalies in the susceptibility and specific heat at 22 K are ascribed to a transition into an antiferromagnetic state. Moderate magnetic fields (<5 kOe) reduce the anomaly in the specific heat shifting it to lower temperatures while higher fields completely suppress this anomaly. A ferromagnetic spin-wave contribution to the specific heat is evidenced at low temperatures. Above 22 K the electron-spin resonance reveals a narrow isotropic line due to ferromagnetic correlations. Below 22 K the spectrum transforms into a broad line associated with the antiferromagnetic correlations and a second line indicating the spin reorientation. The conductivity shows a metal-insulator like transition below 80 K and a maximum near $T_N$. The strong bond frustration due to competing exchange interactions between the Cr ions is suggested to be responsible for the observed magnetic and specific heat anomalies.

**Acknowledgements**

This work was supported by the Deutsche Forschungsgemeinschaft via the Sonderforschungsbereich 484 (Augsburg) and partly by BMBF via VDI/EKM, FKZ 13N6917-B. The support of US CRDF and MRDA via grant MOP2-3050 is gratefully acknowledged.

**Figure captions**

FIG. 1 (color online). X-ray diffraction profiles of $HgCr_2S_4$ poly- (lower frame) and single crystals (upper frame). The measured intensities (open circles) are compared with the calculated profile using Rietveld refinement (solid line). Bragg positions of the normal cubic spinel structure are indicated by vertical bars and the difference pattern by the lower thin solid line. The inset shows the measured intensities of the (511) and (440) reflections of the single (full circles) and polycrystalline (open circles) samples. Scaling of the (511) reflection to the same intensity results in an increase/decrease of the (440) reflection for the polycrystalline/single crystalline sample, respectively.

FIG. 2. Temperature dependences of the magnetization M at different fields for $HgCr_2S_4$ polycrystals (closed symbols) and single crystals (open symbols). Inset: The temperature dependence of the reciprocal susceptibility $\chi^{-1}$ for the same samples measured in a field of 10 kOe. The dashed line indicates a Curie-Weiss behavior of $\chi^{-1}$.

FIG. 3 (color online). Field dependent magnetization curves at different temperatures for $HgCr_2S_4$ single crystals (SC): a) for fields applied along <001> direction; b) for <111> and <001> directions at 2 K on increasing and decreasing field showing hysteresis in low fields and anisotropy in saturation fields. The magnetization curves in the ferromagnetic-like state at 30 K demonstrate similar demagnetization factors of the samples.

FIG. 4 (color online). Temperature dependence of the real part of the susceptibility $\chi'$ at two different external DC magnetic fields for single crystalline (SC) $HgCr_2S_4$ (open symbols) compared to the DC susceptibility measured in a field of 100 Oe (close symbols).

FIG. 5 (color online). Temperature dependences of the specific heat of $HgCr_2S_4$ plotted as $C_p/T$ both for polycrystal (PC, open symbol) and single crystal (SC, close symbol). The inset represents $C_p/T$ *vs.* T at around $T_N$ for the single crystal for different applied fields up to 5 kOe on an enlarged scale.

FIG. 6 (color online): a) Temperature dependence of the specific heat in zero field and for external magnetic fields above 10 kOe. b) Difference between $(C_0-C_{10\ kOe})/T$ versus temperature for poly- (PC) and single crystalline (SC) samples. Inset: Dependences of $C/T^{3/2}$ *vs.* T$^{3/2}$ for different applied fields at low temperatures used to estimate the magnon and phonon contributions. Dashed line shows a fit to the data at zero field.

FIG. 7 (color online). Electron-spin resonance spectra at different temperatures for thin HgCr$_2$S$_4$ single crystalline disc with (110) plane orientation. The static magnetic field is applied along the <001> axis. The solid lines are fits by Lorentz curves, one above and two below 32 K. The arrows mark the spin-reorientation. The numbers on the right indicate the relative amplification factors of the ESR intensity.

FIG. 8. Temperature dependences of the resonance field H$_{res}$ (a) and linewidth ΔH (b) for thin HgCr$_2$S$_4$ single crystalline disc with (110) plane orientation. Open circles indicate the main resonance in the paramagnetic (PM), ferromagnetic (FM) and antiferromagnetic (AFM) states. The solid triangles mark the resonance feature at the spin-reorientation (SR). In the dashed area the two lines overlap and, hence, are difficult to be evaluated separately.

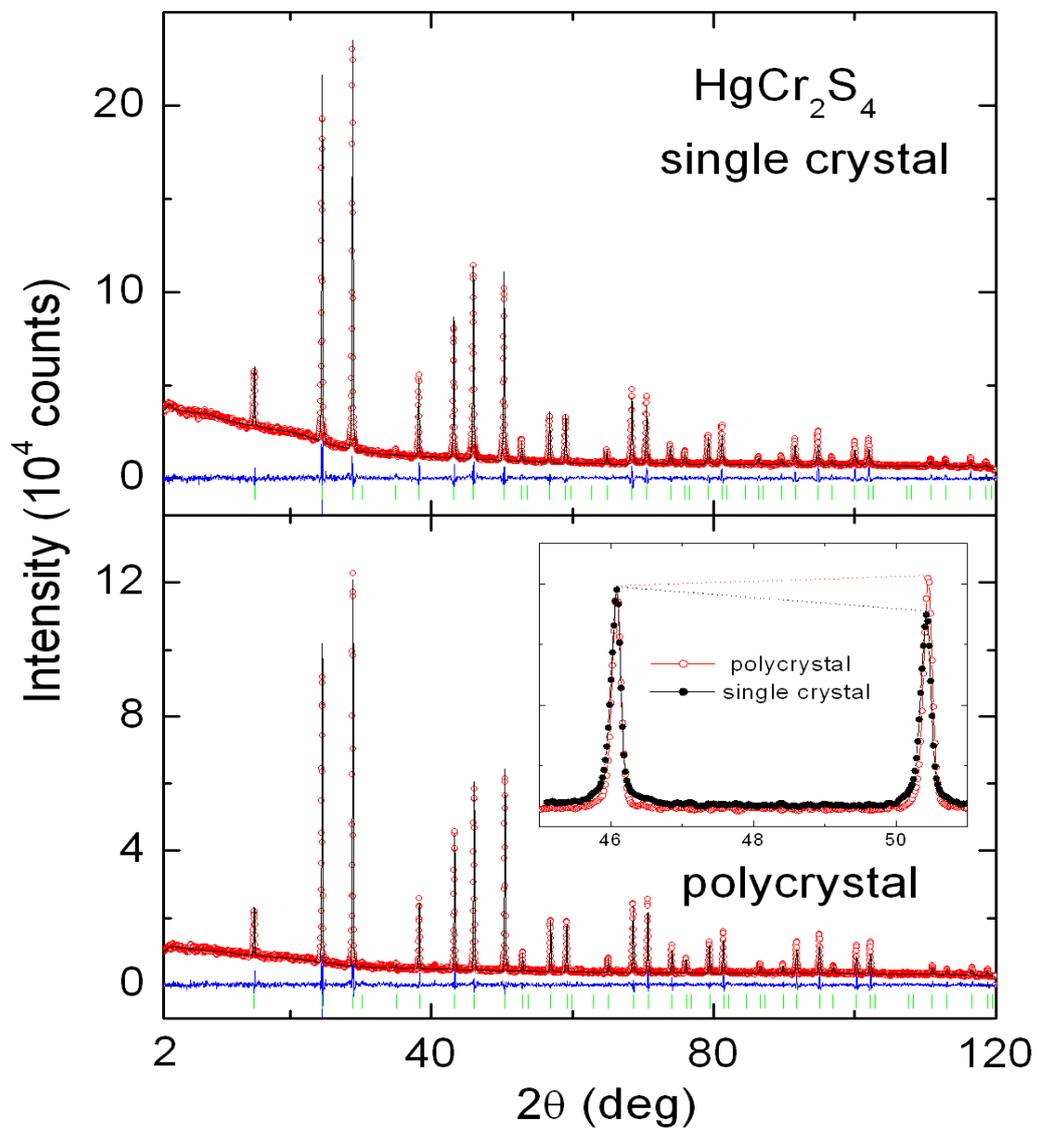

Fig. 1

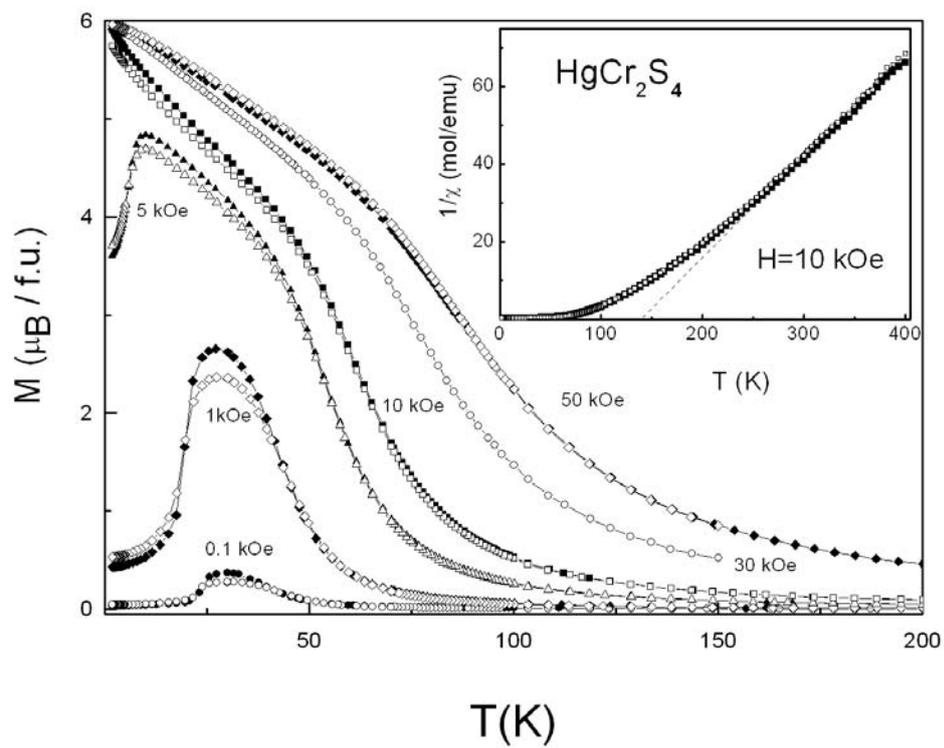

Fig. 2

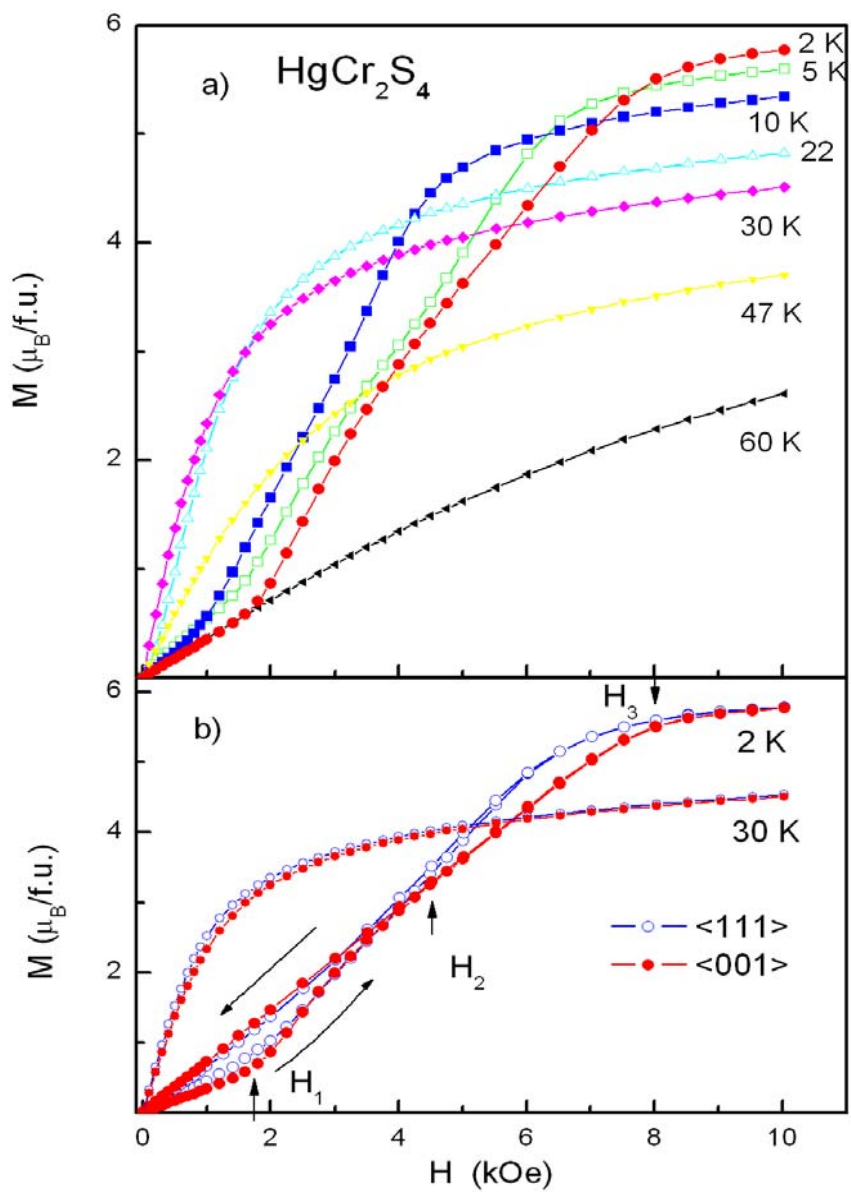

Fig. 3

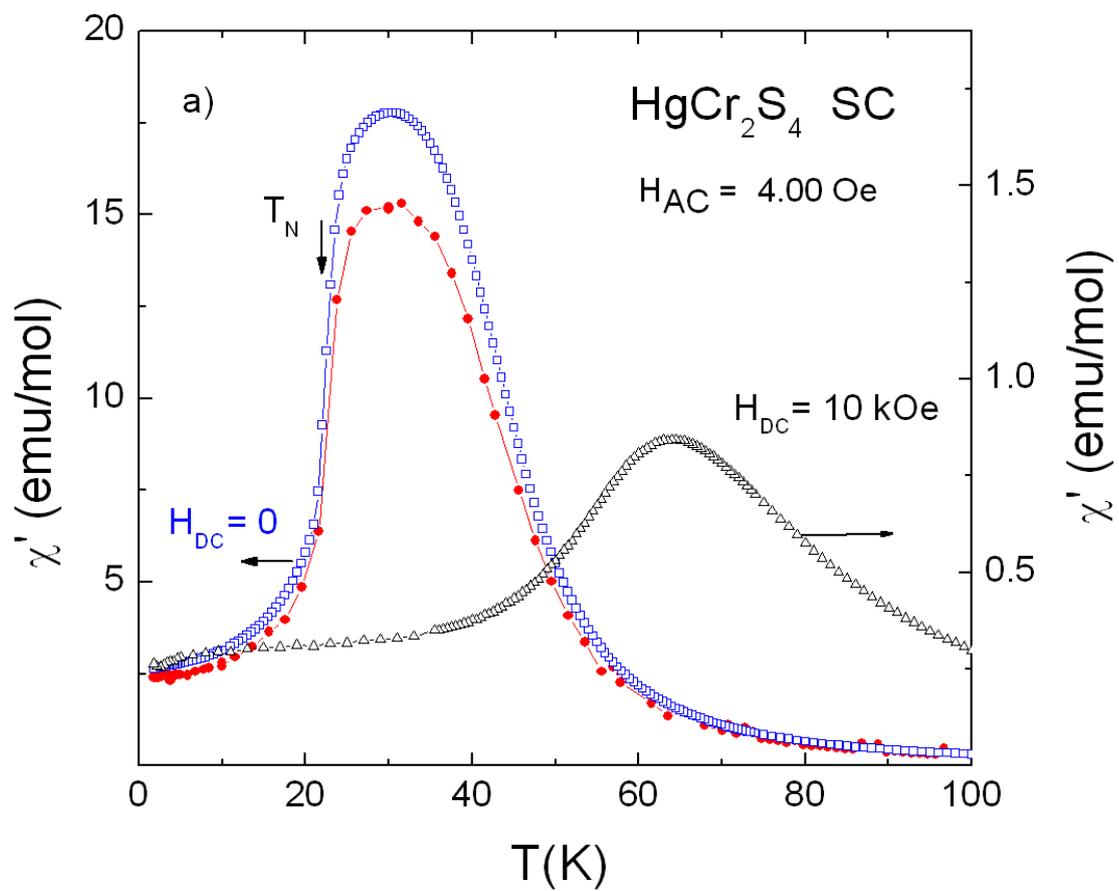

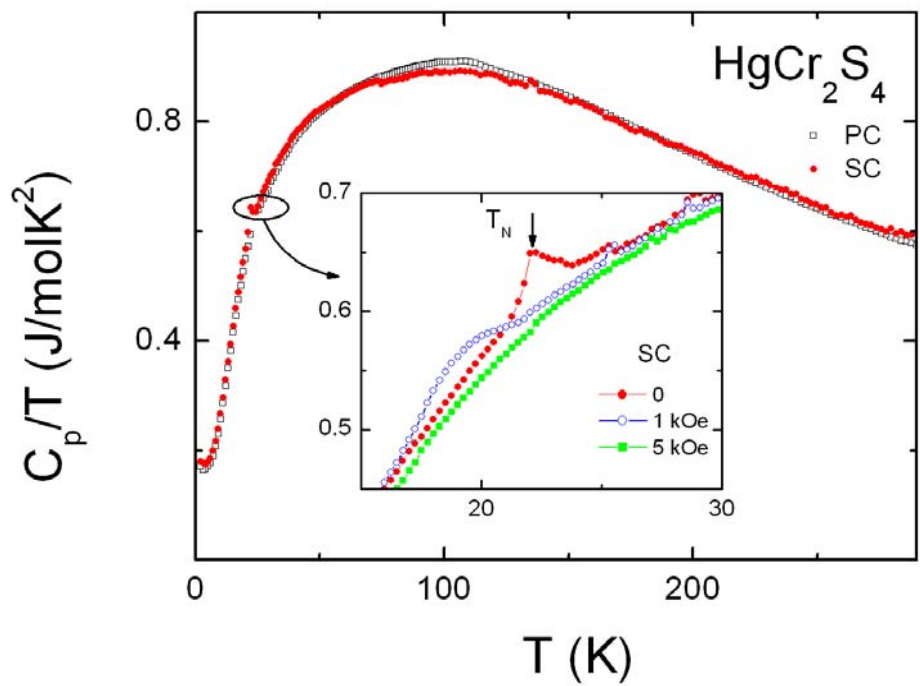

Fig. 5

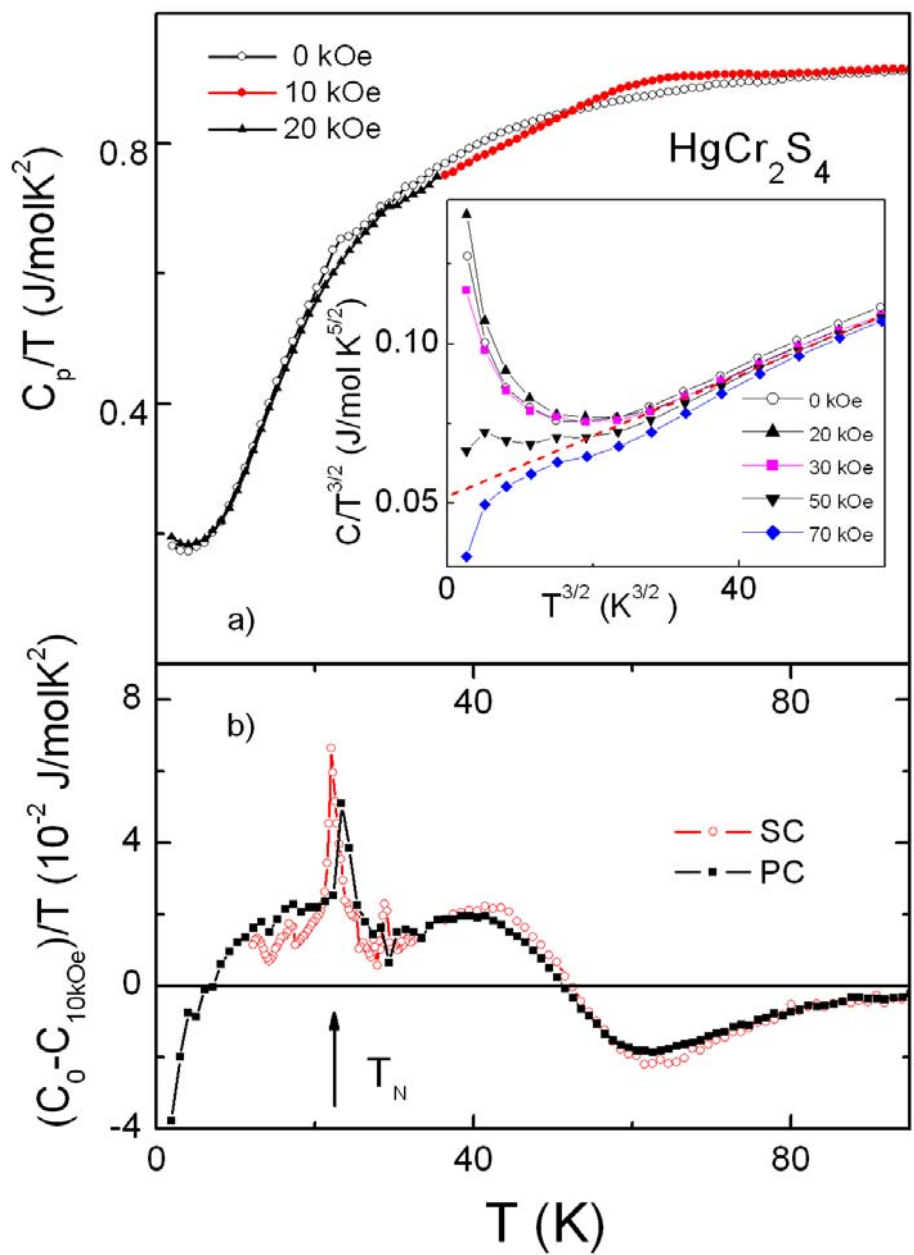

Fig. 6

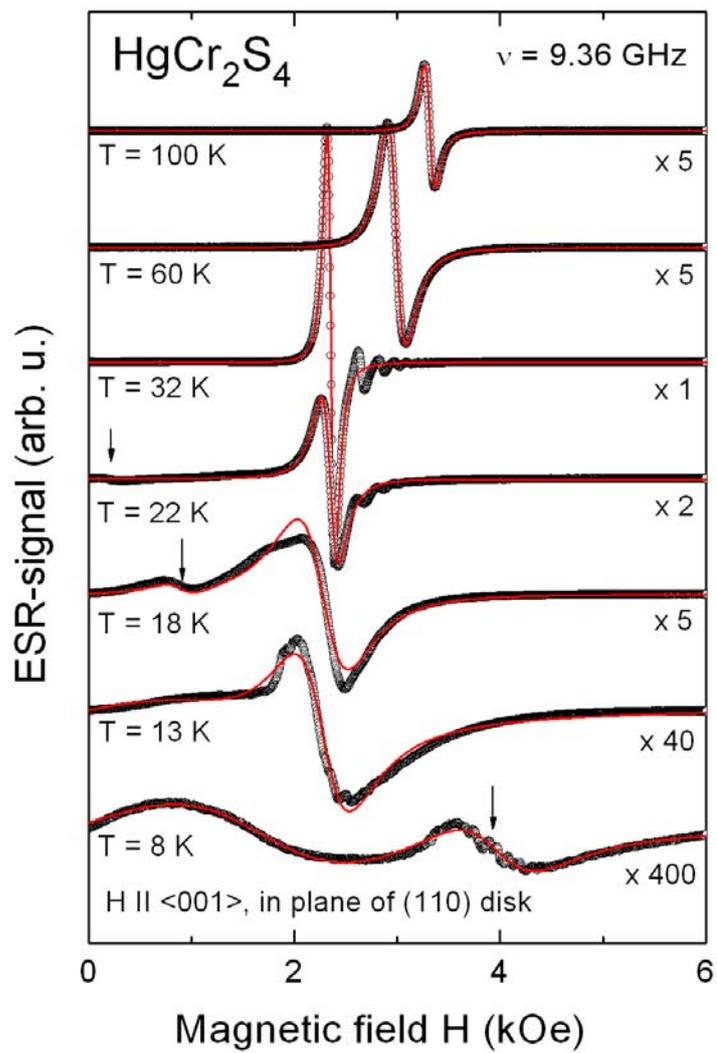

Fig. 7

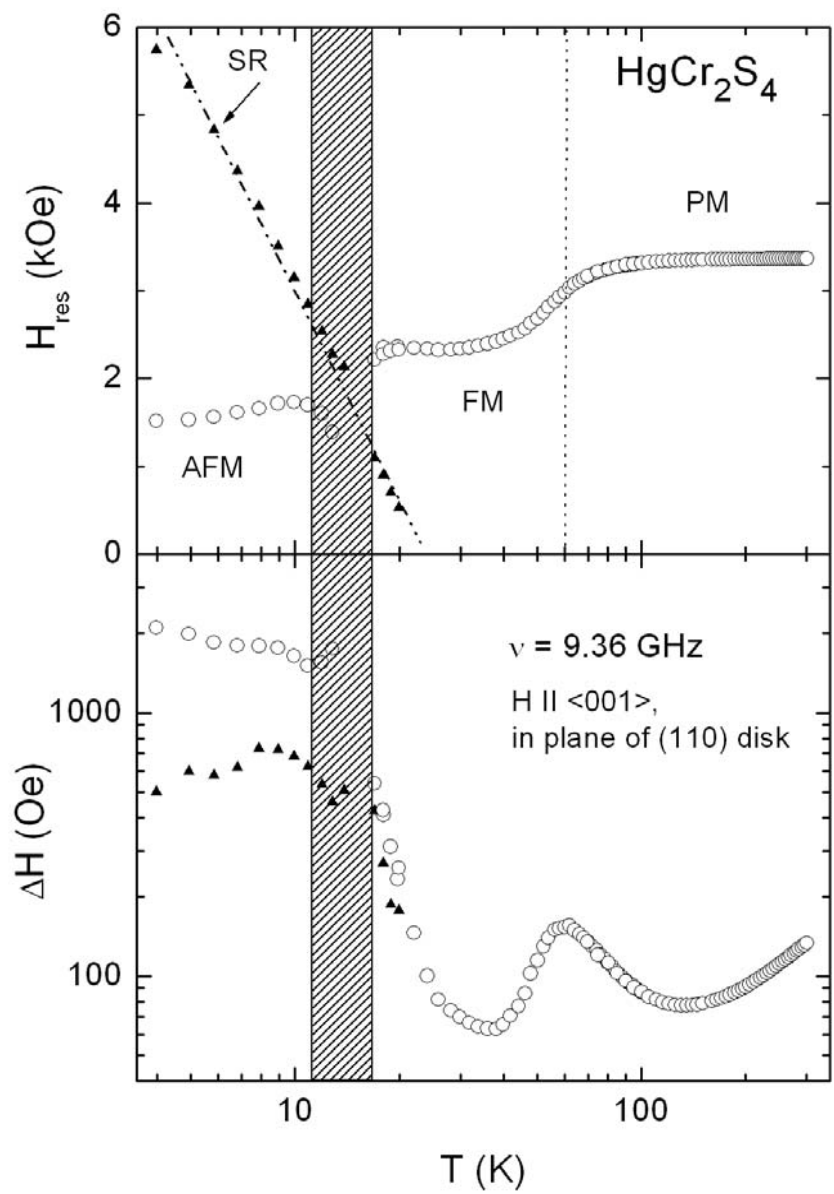

Fig. 8